\documentclass[12pt]{article}
\usepackage{amsmath,amssymb,amsfonts,graphicx}

\makeatletter
\@addtoreset{equation}{section}
\makeatother

\begin{document}
\title{Fractional electric charge of a magnetic vortex at nonzero temperature}
\author{Yurii A. Sitenko\thanks{E-mail:
yusitenko@bitp.kiev.ua}\\
\it \small Bogolyubov Institute for Theoretical Physics,
\it \small National Academy of Sciences of Ukraine,\\
\it \small 14-b Metrologichna str., Kyiv 03143, Ukraine\\
\phantom{11111111111}\\
Volodymyr M. Gorkavenko
\thanks{E-mail: gorka@univ.kiev.ua}\\
\it \small Department of Physics, Taras Shevchenko National University of Kyiv,\\
\it \small 6 Academician Glushkov ave., Kyiv 03680, Ukraine\\
\phantom{11111111111}}
\date{}
\maketitle

\begin{abstract}
An ideal gas of twodimensional Dirac fermions in the background of
a pointlike magnetic vortex with arbitrary flux is considered. We
find that this system acquires fractional electric charge at
finite temperatures and determine the functional dependence of the
thermal average and quadratic fluctuation of the charge on the
temperature, the vortex flux, and the continuous parameter of the
boundary condition at the location of the vortex.
\end{abstract}

\renewcommand{\thesection}{\Roman{section}}
\renewcommand{\theequation}{\arabic{section}.\arabic{equation}}

%%%%%%%%%%%%%%%%%%%%%%%    SECTION 1. INTRODUCTION

\section{Introduction}
Spontaneous breakdown of continuous symmetries can give rise to
topological defects (texture solitons) with rather interesting
properties. A topological defect in threedimensional space, which
is characterized by the nontrivial second homotopy group, is known
as a magnetic monopole \cite{Ho,Pol}, see also genuine
Ref.\cite{Dir}. Vacuum fluctuations of quantized Dirac fields
result in the monopole becoming a CP symmetry violating dyon, i.e.
acquiring nonzero (and fractional) electric charge
\cite{Wit,Gro,Yam}. More recently the effect of thermal
fluctuations of quantized Dirac fields in the presence of the
monopole has been considered, yielding the temperature dependence
of the induced charge \cite{Cor,Gol,Du}.

A topological defect in twodimensional space, which is
characterized by the nontrivial first homotopy group, is a cross
section of the Abrikosov-Nielsen-Olesen magnetic vortex \cite{Abr,
Nie}. The vortex defect is described in terms of a spin-0 field
which condenses and a spin-1 field corresponding to the
spontaneously broken gauge group; the former is coupled to the
latter in the minimal way with constant $e_{cond}$.
Single-valuedness of the condensate field and finiteness of the
vortex energy imply that the vortex flux is related to $e_{cond}$:
\begin{equation}\label{int1}
\Phi=\frac1{2\pi}\oint d
\textbf{x}\,\textbf{V}(\textbf{x})=\frac1{e_{cond}}\,,
\end{equation}
where $\textbf{V}(\textbf{x})$ is the vector potential of the
spin-1 field, and the integral is over a path enclosing once the
vortex tube. The quantized fermion field is coupled minimally to
the spin-1 field with constant $e$ - the elementary charge; thus,
quantum effects depend on the value of $e\Phi$. The case of
$e_{cond}=2e$
$(e\Phi=1/2)$ is realized in ordinary Bardeen-Cooper-Schrieffer
superconductors where the Cooper pair field condenses and, in
addition, there are normal electron (pair-breaking) excitations.
It remains still to be elucidated, whether other values of $e\Phi$
are realized in nature, although there are claims that vortices
with fractional $e\Phi\neq 1/2$ exist in chiral superfluids and
chiral and two-gap superconductors \cite{Vol, BaE}.

The aim of the present paper is to consider the effect of thermal
fluctuations of quantized Dirac fields
\footnote{This may be relevant for various particle physics models
with applications ranging from early Universe cosmology to hot
nuclear matter phenomenology, and even for condensed matter
models, because effectively quasirelativistic fermions arise, in
particular, in $d$-wave type II superconductors (see, e.g.,
Refs.\cite{Dur,Vis,Fra} )}
%%%%%%%%%%%%%%%%%%%%%%%%%%%%%%%%%%%%%%%%%%%%%%%%%%%%%%%%%%
in the presence of the vortex defect with arbitrary value of
$e\Phi$, which results in the vortex acquiring fractional
electric charge; the zero-temperature effect was considered
earlier \cite{Si0,Si6,Si7}. Since continuous symmetry is not
spontaneously broken at the core of the defect, it seems
reasonable to exclude the region of the defect and to impose a
boundary condition for quantized fields at the edge of this
region. Thus, quantum effects depend both on
$e\Phi$ and real continuous quantity $\Theta$ which parameterizes
the most general varieties of boundary conditions (for more
details see next Section). This setup should not be confused with
the setup  when fermions are quantized in the presence of an
extensive magnetic field with finite flux and the region of the
nonvanishing field strength is not excluded. The induced charge in
the latter case was considered in Refs.\cite{NiS, Red, Jac4}(zero
temperature) and Refs.\cite{Niemi, Bab, Po}(nonzero temperature),
and we shall compare the results of both setups in Section V.

The operator of the second-quantized fermion field in a static
background can be presented in the form
\begin{equation}\label{int2}
\Psi(\textbf{x},
t)=\sum\hspace{-1.7em}\int\limits_{E>0}e^{-iEt}\langle\textbf{x}|E,
\lambda\rangle a_{E\lambda}+
\sum\hspace{-1.7em}\int\limits_{E<0}e^{-iEt}\langle\textbf{x}|E,
\lambda\rangle b^+_{E\lambda}\,,
\end{equation}
where $a^+_{E\lambda}$ and $a_{E\lambda}$ $(b^+_{E\lambda}$ and
$b_{E\lambda})$ are the fermion (antifermion) creation and
destruction  operators satisfying anticommutation relations,
\begin{equation}\label{int3}
\left[a_{E\lambda},a^+_{E'\lambda'}\right]_+=
\left[b_{E\lambda},b^+_{E'\lambda'}\right]_+= \langle
E,\lambda|E',\lambda'\rangle\,,
\end{equation}
and $\langle \textbf{x}|E,\lambda\rangle$ is the solution to the
stationary Dirac equation,
\begin{equation}\label{int4}
H\langle \textbf{x}|E,\lambda\rangle=E\langle
\textbf{x}|E,\lambda\rangle\,,
\end{equation}
$H$ is the Dirac Hamiltonian, $E$ is the energy and $\lambda$ is
the set of other parameters (quantum numbers) specifying a state;
symbol ${\displaystyle\sum\hspace{-1.4em}\int\,}$ means the
summation over discrete and the integration (with a certain
measure) over continuous values of all quantum numbers.
Conventionally, the operators of dynamical invariants are defined
as bilinears of the fermion field operators, and, thus,
comprizing:

\noindent the energy operator (temporal component of the
energy-momentum vector),
\begin{equation}\label{int5}
\hat P^0=\frac i4 \int d^d x
\left(\left[\Psi^+,\partial_t\Psi\right]_--\left[\partial_t\Psi^+,\Psi\right]_-\right)\,,
\end{equation}
and the fermion number operator,
\begin{equation}\label{int6}
 \hat N =\frac12 \int d^d x \left[\Psi^+,\Psi\right]_-\,,
\end{equation}
where $d$ is the space dimension. Operators (\ref{int5}) and
(\ref{int6}) commute and are thus diagonal in the fermion and
antifermion creation and destruction operators.

The thermal average of the fermion number operator over the
canonical ensemble is defined as (see, e.g., Ref.\cite{Das})
\begin{equation}\label{int7}
\langle \hat N\rangle=\frac{Sp \, \hat N\, {\rm{exp}}(-\beta\hat
P^0)}{Sp\,\,{\rm{exp}}(-\beta\hat
P^0)}\,,\qquad\beta=(k_BT)^{-1}\,,
\end{equation}
where $T$ is the equilibrium temperature, $k_B$ is the Boltzmann constant, and
$Sp$ is the trace or the sum over the expectation values in the Fock state
basis created by operators in Eq.(\ref{int3}). Appropriately, the electric
charge of the quantum fermionic system in thermal equilibrium is given by
expression
\begin{equation}\label{int8}
Q(T)\equiv e\langle \hat N\rangle=-\frac
e2\int\limits_{-\infty}^\infty dE\,\tau(E)\tanh\left(\frac12\beta
E\right)\,,
\end{equation}
where the last equality is obtained by transforming the right hand
side of Eq.(\ref{int7}) into an integral over the spectrum of the
Dirac Hamiltonian (see, e.g., Ref.\cite{Niemi}), and the spectral
density of the Dirac Hamiltonian (or density of states) is
\begin{equation}\label{int9}
\tau(E)=\frac1\pi Im\,\,Tr\frac1{H-E-i0}\,\,,
\end{equation}
where $Tr$ is the trace of an integro-differential operator in
functional space: $Tr U=$
\mbox{$=\int d^dx\,tr\langle\textbf{x}|U|\textbf{x}\rangle$}; $tr$ denotes the
trace over spinor indices only;  note that the functional trace
should be regularized and renormalized by subtraction, if
necessary.
{\sloppy

}

Similarly, one gets expression for the quadratic fluctuation of
the electric charge:
\begin{equation}\label{int10}
\Delta^2_{Q(T)}\equiv e^2\left[\langle \hat N^2\rangle-\left(\langle
\hat N\rangle\right)^2\right]
=\frac{e^2}{4}\int\limits_{-\infty}^\infty
dE\,\frac{\tau(E)}{\cosh^2\left(\frac12\beta E\right)}\,.
\end{equation}
Evidently, if the quadratic fluctuation becomes nonvanishing,
then the corresponding dynamical invariant ceases to be a sharp
quantum observable.

In the present paper we shall find electric charge (\ref{int8})
and its fluctuation (\ref{int10}) in the $d=2$ quantum fermionic
system in the background of a single static topological defect
which is a twodimensional cross section of the magnetic vortex.

%%%%%%%%%%%%%%%%%%%%%%%%%%          SECTION 2

\section{Self-adjointness of the Dirac Hamiltonian
 in the background of the pointlike vortex defect}
 The Dirac Hamiltonian in external magnetic field takes form
\begin{equation}\label{b1}
H=-i\gamma^0{\mbox{\boldmath $\gamma$}} \left[{\mbox{\boldmath
$\partial $}}-ie\textbf{V}(\textbf{x})\right]+\gamma^0m\,.
\end{equation}
In $2+1$-dimensional space-time $(\textbf{x}, t)=(x^1,x^2,t)$,
the Clifford algebra has two inequivalent irreducible
representations which can be differed in the following way:
\begin{equation}\label{b2}
i\gamma^0\gamma^1\gamma^2=s, \qquad s=\pm1\,.
\end{equation}
Choosing the $\gamma^0$ matrix in the diagonal form,
\begin{equation}\label{b3}
\gamma^0=\sigma_3\,,
\end{equation}
one gets
\begin{equation}\label{b4}
\gamma^1=e^{\frac i2 \sigma_3\chi_s}i\sigma_1e^{\frac{-i}2
\sigma_3\chi_s}\,,\qquad \gamma^2=e^{\frac i2
\sigma_3\chi_s}is\sigma_2e^{\frac{-i}2 \sigma_3\chi_s}\,,
\end{equation}
where $\sigma_1$, $\sigma_2$, and $\sigma_3$ are the Pauli
matrices, and $\chi_1$ and $\chi_{-1}$ are the parameters varying
in interval $0<\chi_s<2\pi$ to go over to the equivalent
representation. Note also that in odd-dimensional space-time the
$m$ parameter in Eq.(\ref{b1}) can take both positive and
negative values; a change of sign of $m$ corresponds to going
over to the inequivalent representation.

A solution to stationary Dirac equation (\ref{int4}) with
Hamiltonian (\ref{b1}) can be presented as
\begin{equation}\label{b5}
\langle\textbf{x}|E,n\rangle=\left(\begin{array}{ll}
f_n(r,E)&e^{in\varphi+i\chi_s}\\
g_n(r,E)&e^{i(n+s)\varphi}\end{array}\right)\,,\qquad n\in
\mathbb Z\,,
\end{equation}
where polar coordinates
$r=\sqrt{{\left(x^1\right)}^2+{\left(x^2\right)}^2}$ and
$\varphi=\arctan(x^2/x^1)$ are introduced, and $\mathbb Z$ is the
set of integer numbers. The magnetic field strength and its flux
in the units of $2\pi$ are given by expressions (compare with
Eq.(\ref{int1})):
\begin{equation}\label{b6}
B(\textbf{x})={\mbox{\boldmath $\partial $}}
\times\textbf{V}(\textbf{x})\,,\qquad \Phi=\frac1{2\pi}\int
d^2x\,B(\textbf{x})\,.
\end{equation}
Since a single defect is considered, a support of the magnetic
field strength is localized in a certain, let it be central,
region of twodimensional space. It is evident that different
functions $B(\textbf{x})$ can give the same value of $\Phi$. In
general, a solution to the Dirac equation in an external magnetic
field depends on the field configuration in two ways: there is a
direct, or local, impact of $B(\textbf{x})$ on the solution at the
same point $\textbf{x}$ (similar to the action of the classical
Lorentz force), and there is an indirect (through vector
potential) influence of the field strength on the behaviour of
the solution in regions out of the field strength's support
(similar to the quantum-mechanical Bohm-Aharonov effect
\cite{Aha}). Namely the latter effects are those which interest
us, since, as it has been already noted in Introduction, the
central region (i.e. the region of the field strength's support)
is excluded, and at its edge a boundary condition is imposed on
the solution to the Dirac equation.

It might be anticipated that the solution in the outer region
depends on the flux rather than the local features of the field
strength in the central region. However, a closer look at the
situation when the central region is not excluded suggests that
the local features of the field strength may influence the
behaviour of the solution out of the central region. Really,
changes of a profile of the field strength influence strongly the
behaviour of the solution in the central region, and this, due to
the continuity and smoothness properties of the solution as a
solution to a differential equation, entails changes in its
behaviour in the outer region. Thus, when the central region is
excluded, our purpose is not to stick to a limited set of
boundary conditions but, instead, to extend this set maximally in
order to cover all possible types of the behaviour of the
solution near the boundary and, perhaps, all plausible profiles
of the field strength in the excluded region.

How to achieve this purpose in general, remains to be a question. However, a
recipe is available under a simplifying assumption that finite size of the
excluded region is neglected: then the most general conditions are those
ensuring self-adjointness of the Dirac Hamiltonian, and they are labelled by
self-adjoint extension parameter $\Theta$ (see, e.g., Ref.\cite{Jac1}).
Although $\Theta$ is physically interpreted as the CP violating vacuum angle in
the $d=3$ case of the monopole defect \cite{Wit, Gro, Yam}, the direct physical
interpretation of $\Theta$ in the $d=2$ case of the vortex defect is yet
lacking. Both in the monopole and vortex cases parameter $\Theta$ is involved
into a condition for just one of the modes of the solution to the Dirac
equation.

When the transverse size of the vortex defect is shrinked to
zero, the magnetic field strength takes form
\begin{equation}\label{b7}
B(\textbf{x})=2\pi\Phi\delta(\textbf{x})\,,
\end{equation}
and the vector potential can be chosen as
\begin{equation}\label{b8}
V^1(\textbf{x})=-\Phi \,r^{-1}\sin\varphi\,,\qquad V^2(\textbf{
x})=\Phi\, r^{-1}\cos\varphi\,.
\end{equation}
Then Dirac Hamiltonian (\ref{b1}) takes form
\begin{equation}\label{b9}
H=-i\gamma^0\gamma^r\partial_r-ir^{-1}\gamma^0\gamma^{\varphi}(\partial_\varphi-ie\Phi)+\gamma^0m\,,
\end{equation}
where
\begin{equation}\label{b10}
\gamma^r=\gamma^1\cos\varphi+\gamma^2\sin\varphi\,,\qquad
\gamma^\varphi=-\gamma^1\sin\varphi+\gamma^2\cos\varphi\,.
\end{equation}
Using explicit form of $\gamma$ matrices (\ref{b3})-(\ref{b4}),
one finds that the Dirac equation in the background of a
pointlike defect is reduced to following set of equations for the
modes of Eq.(\ref{b5}):
\begin{equation}\label{b11}
\left(\begin{array}{cc}
m&\partial_r+s(n-e\Phi+s)r^{-1}\\
-\partial_r+s(n-e\Phi)r^{-1}&-m\end{array}\right)
\left(\begin{array}{c} f_n\\g_n\end{array}\right)= E
\left(\begin{array}{c} f_n\\g_n\end{array}\right)\,.
\end{equation}
Partial Hamiltonians are essentially self-adjoint for all $n$,
with the exception of $n=n_0$, where
\begin{equation}\label{b12}
n_0=[\![e\Phi]\!]+\frac12-\frac12s\,,
\end{equation}
$[\![u]\!]$ is the integer part of quantity $u$ (i.e., the
largest integer which is less than or equal to $u$).
Correspondingly, the modes with $n\neq n_0$ are regular at $r=0$
(i.e. at the location of the defect). The partial Hamiltonian at
$n=n_0$ requires a self-adjoint extension according to the
Weyl-von Neumann theory of self-adjoint operators (see, e.g.,
Ref.\cite{Alb}), which upon implementation yields following
condition for the corresponding mode \cite{Ger, Si6, Si7}:
\begin{equation}\label{b13}
\cos\left(s\frac\Theta2+\frac\pi4\right)\lim_{r\rightarrow0}(|m|r)^Ff_{n_0}=
-{\rm{sgn}}(m)\sin\left(s\frac\Theta2+\frac\pi4\right)\lim_{r\rightarrow0}(|m|r)^{1-F}g_{n_0}\,,
\end{equation}
where
\begin{equation*}
{\rm{sgn}}(u)= \left\{\begin{array}{cc} 1\,,& u>0\\ -1\,,&
u<0\end{array}\right\}\,,
\end{equation*}
$\Theta$ is the self-adjoint extension parameter, and
\begin{equation}\label{b14}
F=s\{\![e\Phi]\!\}+\frac12-\frac12s\,,
\end{equation}
$\{\![u]\!\}=u-[\![u]\!]$ is the fractional part of quantity $u$,
$0\leq \{\![u]\!\} <1$; note here that Eq.(\ref{b13}) implies
that $0<F<1$, since in the case of
${\displaystyle F=\frac12-\frac12s}$ both
$f_{n_0}$ and $g_{n_0}$ obey the condition of regularity at
$r\rightarrow0$. Note also that Eq.(\ref{b13}) is periodic in
$\Theta$ with period $2\pi$.

So far solutions corresponding to the continuous spectrum,
$|E|>|m|$, are concerned, that obey the "orthonormality" condition
\begin{equation}\label{b15}
\int d^2 x\langle E, n|\textbf{x}\rangle\langle\textbf{x}|E',
n'\rangle=\frac{\delta(E-E')}{\sqrt{|EE'|}}\delta_{nn'}\,.
\end{equation}
Owing to Eq.(\ref{b13}), an additional solution corresponding to
the bound state with energy $E=E_{BS}$, $|E_{BS}|<|m|$, appears at
$\cos \Theta<0$, that obeys usual normalization condition
\begin{equation}\label{b16}
\int d^2 x\langle E_{BS}, n_0|\textbf{x}\rangle\langle\textbf{
x}|E_{BS}, n_0 \rangle=1\,.
\end{equation}
Its energy is determined as a real root of algebraic equation
\begin{equation}\label{b17}
\frac{(1+m^{-1}E_{BS})^{1-F}}{(1-m^{-1}E_{BS})^F}=-A\,,
\end{equation}
where
\begin{equation}\label{b18}
A=2^{1-2F}\frac{\Gamma(1-F)}{\Gamma(F)}\tan\left(s\frac\Theta2+\frac\pi4\right)\,,
\end{equation}
$\Gamma(u)$ is the Euler gamma function. The bound state energy is
zero, $E_{BS}=0$, at $A=-1$; otherwise, we get
\begin{equation}\label{b19}
{\rm{sgn}}(E_{BS})=\frac12{\rm{sgn}}(m)[{\rm{sgn}}(1+A^{-1})-{\rm{sgn}}(1+A)]\,.
\end{equation}
At $\cos\Theta>0$ $(A>0)$ the right hand side of Eq.(\ref{b19})
turns to zero, which corresponds to the absence of bound state in
this case.

%%%%%%%%%%%%%%%%%%%%%%%%%%          SECTION 3

\section{Resolvent and spectral density}
The kernel of the resolvent (the Green's function) of the Dirac
Hamiltonian in the coordinate representation is defined as
\begin{equation}\label{c1}
G^\omega(r,\varphi;r',\varphi')=\langle
r,\varphi|(H-\omega)^{-1}|r',\varphi' \rangle,
\end{equation}
where $\omega$ is a complex parameter with dimension of energy.
The expansion of Eq.(\ref{c1}) in modes takes form
\begin{equation}\label{c2}
G^\omega(r,\varphi;r',\varphi')=\frac1{2\pi}\sum_{n\in \mathbb Z}
e^{in(\varphi-\varphi')} \left(\begin{array}{cc} a_n(r;r')&
d_n(r;r')e^{-i(s\varphi'-\chi_s)}\\
b_n(r;r')e^{i(s\varphi-\chi_s)} &
c_n(r;r')e^{is(\varphi-\varphi')}\end{array}\right).
\end{equation}
In the case of $H$ given by Eq.(\ref{b9}) radial components of
$G^\omega(r,\varphi;r',\varphi')$ (\ref{c2}) satisfy equations
(compare with Eq.(\ref{b11})):
\begin{multline}\label{c3}
\left(\begin{array}{cc}
-(\omega-m)&\partial_r+s(n-e\Phi+s)r^{-1}\\
-\partial_r+s(n-e\Phi)r^{-1}&-(\omega+m)\end{array}\right)
\left(\begin{array}{cc}
a_n(r;r')& d_n(r;r')\\
b_n(r;r') & c_n(r;r')\end{array}\right)=\\
 =\left(\begin{array}{cc}
-(\omega-m)&\partial_{r'}+s(n-e\Phi+s){r'}^{-1}\\
-\partial_{r'}+s(n-e\Phi){r'}^{-1}&-(\omega+m)\end{array}\right)
\left(\begin{array}{cc}
a_n(r;r')& b_n(r;r')\\ d_n(r;r') & c_n(r;r')\end{array}\right)=\\
=\frac{\delta(r-r')}{\sqrt{rr'}} \left(\begin{array}{cc} 1& 0\\ 0
& 1\end{array}\right).
\end{multline}
Off-diagonal radial components are expressed through the diagonal
ones:
\begin{multline}\label{c4}
b_n(r;r')=(\omega+m)^{-1}\left[-\partial_r+s(n-e\Phi)r^{-1}\right]a_n(r;r')=\\
=(\omega-m)^{-1}\left[\partial_{r'}+s(n-e\Phi+s){r'}^{-1}\right]c_n(r;r'),
\end{multline}
\begin{multline}\label{c5}
d_n(r;r')=(\omega-m)^{-1}\left[\partial_r+s(n-e\Phi+s)r^{-1}\right]c_n(r;r')=\\
=(\omega+m)^{-1}\left[-\partial_{r'}+s(n-e\Phi){r'}^{-1}\right]a_n(r;r').
\end{multline}
In Appendix A we determine the diagonal radial components which
can be presented in the following way,

type 1 $(l=s(n-n_0)>0)$:
\begin{equation}\label{c6}
a_n(r;r')=\frac{i\pi}2(\omega+m)\left[\theta(r-r')H^{(1)}_{l-F}(kr)J_{l-F}(kr')+
\theta(r'-r)J_{l-F}(kr)H^{(1)}_{l-F}(kr')\right],
\end{equation}
\begin{equation}\label{c7}
c_n(r;r')=\frac{i\pi}2(\omega-m)\left[\theta(r-r')H^{(1)}_{l+1-F}(kr)J_{l+1-F}(kr')+
\theta(r'-r)J_{l+1-F}(kr)H^{(1)}_{l+1-F}(kr')\right];
\end{equation}

type 2 $(l'=-s(n-n_0)>0)$:
\begin{equation}\label{c8}
a_n(r;r')=\frac{i\pi}2(\omega+m)\left[\theta(r-r')H^{(1)}_{l'+F}(kr)J_{l'+F}(kr')+
\theta(r'-r)J_{l'+F}(kr)H^{(1)}_{l'+F}(kr')\right],
\end{equation}
\begin{equation}\label{c9}
c_n(r;r')=\frac{i\pi}2(\omega-m)\left[\theta(r-r')H^{(1)}_{l'-1+F}(kr)J_{l'-1+F}(kr')+
\theta(r'-r)J_{l'-1+F}(kr)H^{(1)}_{l'-1+F}(kr')\right];
\end{equation}

type 3 $(n=n_0)$:
\begin{multline}\label{c10}
a_{n_0}(r;r')=\frac{i\pi}2\frac{\omega+m}{\sin\nu_\omega+\cos\nu_\omega
e^{iF\pi}}\left\{\theta(r-r')H^{(1)}_{-F}(kr) [\sin\nu_\omega
J_{-F}(kr')+\cos\nu_\omega J_{F}(kr')]+\right.\\
+\left.\theta(r'-r)[\sin\nu_\omega J_{-F}(kr)+\cos\nu_\omega
J_{F}(kr)]H^{(1)}_{-F}(kr') \right\},
\end{multline}
\begin{multline}\label{c11}
c_{n_0}(r;r')=\frac{i\pi}2\frac{\omega-m}{\sin\nu_\omega+\cos\nu_\omega
e^{iF\pi}}\left\{\theta(r-r')H^{(1)}_{1-F}(kr) [\sin\nu_\omega
J_{1-F}(kr')-\cos\nu_\omega J_{-1+F}(kr')]+\right.\\
+\left.\theta(r'-r)[\sin\nu_\omega J_{1-F}(kr)-\cos\nu_\omega
J_{-1+F}(kr)]H^{(1)}_{1-F}(kr') \right\}.
\end{multline}
Here $k=\sqrt{\omega^2-m^2}$, ${\displaystyle\theta(u)=\frac12
[1+{\rm{sgn}}(u)]}$, $J_\lambda(u)$ is the Bessel function of
order
$\lambda$, $H^{(1)}_\lambda(u)$ is the first-kind Hankel function
of order $\lambda$, and
\begin{equation}\label{c12}
\tan\nu_\omega=\frac{k^{2F}}{\omega+m}\,{\rm{sgn}}(m)|m|^{1-2F}A\,,
\end{equation}
where $A$ is given by Eq.(\ref{b18}). Note that the type 1 and
type 2 components are regular at $r=0$ (or $r'=0$), whereas the
type 3 components are irregular at $r=0$ (or $r'=0$), satisfying
conditions (compare with Eq.(\ref{b13})):
\begin{equation}\label{c13}
\cos\left(s\frac\Theta2+\frac\pi4\right)
\lim\limits_{r\rightarrow0}(|m|r)^Fa_{n_0}(r;r')= -{\rm
sgn}(m)\sin\left(s\frac\Theta2+\frac\pi4\right)
\lim\limits_{r\rightarrow0}(|m|r)^{1-F}b_{n_0}(r;r'),
\end{equation}
\begin{equation}\label{c13b}
\cos\left(s\frac\Theta2+\frac\pi4\right)
\lim\limits_{r\rightarrow0}(|m|r)^Fd_{n_0}(r;r')= -{\rm
sgn}(m)\sin\left(s\frac\Theta2+\frac\pi4\right)
\lim\limits_{r\rightarrow0}(|m|r)^{1-F}c_{n_0}(r;r'),
\end{equation}
and the  ones at $r'\rightarrow0$, which are obtained from
Eqs.(\ref{c13}) and (\ref{c13b}) by interchange
$b_{n_0}\longleftrightarrow d_{n_{0}}$.

Taking $r'>r$ for definiteness, we get relations
\begin{multline}\label{c14}
\int\limits_0^{2\pi}d\varphi\,\,tr\,G^\omega(r,\varphi;r',\varphi)=
\sum_{n\in\,\mathbb Z}\left[a_n(r;r')+c_n(r;r')\right]=\\
=\sum_{\begin{array}{c}\raisebox{0.3em}{${\scriptstyle l\in
\,\mathbb Z}$}\vspace{-0.7em}
\\{\scriptstyle l\geq 1}\end{array}}
\left[(\omega+m)I_{l-F}(-ikr)K_{l-F}(-ikr')+
(\omega-m)I_{l+1-F}(-ikr)K_{l+1-F}(-ikr')\right]+\\
+\sum_{\begin{array}{c}\raisebox{0.3em}{${\scriptstyle l'\in
\,\mathbb Z}$}\vspace{-0.7em}
\\{\scriptstyle l'\geq 1}\end{array}}
\left[(\omega+m)I_{l'+F}(-ikr)K_{l'+F}(-ikr')+
(\omega-m)I_{l'-1+F}(-ikr)K_{l'-1+F}(-ikr')\right]+\\
+(\omega+m)I_{F}(-ikr)K_{F}(-ikr')+(\omega-m)I_{1-F}(-ikr)K_{1-F}(-ikr')
+\frac{2\sin(F\pi)}{\pi\left(\tan\nu_\omega+e^{iF\pi}\right)}\times\\
\times\left[(\omega+m)\tan\nu_\omega
K_{F}(-ikr)K_{F}(-ikr')+(\omega-m)e^{iF\pi}K_{1-F}(-ikr)K_{1-F}(-ikr')\right],
\end{multline}
and
\begin{multline}\label{c15}
\int\limits_0^{2\pi}\left.\vphantom{\int}d\varphi\,\,tr\,G^\omega(r,\varphi;r',\varphi)\right|_{e\Phi=0}=
\sum_{n\in\,\mathbb Z}\left.\vphantom{\sum}\left[a_n(r;r')+c_n(r;r')\right]\right|_{e\Phi=0}=\\
=2\omega\sum_{n\in\,\mathbb Z}I_n(-ikr)K_n(-ikr')\,,
\end{multline}
where $I_\lambda(u)$ is the modified Bessel function of order
$\lambda$, and
\begin{equation*}
K_\lambda(u)=\frac{\pi}{2\sin(\lambda\pi)}\left[I_{-\lambda}(u)-I_{\lambda}(u)\right]
\end{equation*}
is the Macdonald function  of order $\lambda$; note that the last
equalities in Eqs.(\ref{c14}) and (\ref{c15}) are obtained under
condition {\em Im}\,$k>0$. Using relations (see, e.g.,
Ref.\cite{Prud})
\begin{equation*}
I_\lambda(\kappa r)K_\lambda(\kappa
r')=\frac12\int\limits_0^\infty\frac{d\,y}{y}
\exp\left(-\frac{\kappa^2rr'}{2y}-\frac{r^2+{r'}^2}{2rr'}y\right)
I_\lambda(y),\quad\mbox{\em Re}\,\kappa^2>0,
\end{equation*}
\begin{equation*}
\sum\limits_{\begin{array}{c}\raisebox{0.3em}{${\scriptstyle l\in
\,\mathbb Z}$}\vspace{-0.7em}
\\{\scriptstyle l\geq 1}\end{array}}
 I_{l+\lambda}(y)=-\frac1{2\lambda}
\left\{e^y\int\limits_0^y du\,e^{-u}I_\lambda(u)-
y\left[I_\lambda(y)+I_{\lambda+1}(y)\right]\right\},
\quad\mbox{\em Re}\,\lambda>-1,
\end{equation*}
we perform summation in Eqs.(\ref{c14}) and (\ref{c15}) and get in
the case of $\mbox{\em Im}\,k>|\mbox{\em Re}\,k|$:
\begin{multline}\label{c16}
\int\limits_0^{2\pi}d\varphi\,\,tr\left[G^\omega(r,\varphi;r',\varphi)-
\left.\vphantom{\int}G^\omega(r,\varphi;r',\varphi)\right|_{e\Phi=0}\right]=
\frac{2\sin(F\pi)}{\pi\left(\tan\nu_\omega+e^{iF\pi}\right)}\times\\
\times\left[(\omega+m)\tan\nu_\omega K_{F}(\kappa r)K_{F}(\kappa
r')+(\omega-m)e^{iF\pi}K_{1-F}(\kappa r)K_{1-F}(\kappa r')\right]-\\
-\frac{\sin(F\pi)}{2\pi F(1-F)}\omega
\int\limits_0^\infty dy\,\exp\left[-\frac{\kappa^2rr'}{2y}-\frac{(r-r')^2}{2rr'}y\right]\times\\
\times\left\{\frac1{y} \int\limits_y^\infty du\,e^{-u}[(1-F)K_F(u)+F\,K_{1-F}(u)]-e^{-y}(2F-1)
[K_F(y)-K_{1-F}(y)]\right\},
\end{multline}
where $\kappa=-ik$. Taking the limit $r'\rightarrow r$ in
Eq.(\ref{c16}) and integrating it over the radial variable, we get
the renormalized (finite) trace of the resolvent operator
\begin{multline}\label{c17}
Tr\,(H-\omega)^{-1}\equiv\int\limits_0^\infty dr\,
r\int\limits_0^{2\pi}d\varphi\,tr\left[G^\omega(r,\varphi;r,\varphi)-
\left.\vphantom{\int}G^\omega(r,\varphi;r,\varphi)\right|_{e\Phi=0}\right]=\\
=\frac1{\omega^2-m^2}\left[\omega\left(\frac{2F-1}{e^{-iF\pi}\tan\nu_\omega+1}-F^2\right)
+m\left(\frac{1}{e^{-iF\pi}\tan\nu_\omega+1}-F\right)\right];
\end{multline}
note that the last result can be continued analytically to the whole complex
$\omega$-plane. Note also that Eq.(\ref{c17}) can be rewritten in an equivalent
form:
\begin{multline}\label{c18}
Tr\,(H-\omega)^{-1}=\\
=\frac1{\omega^2-m^2}\left\{\omega\left[\frac{2F-1}{e^{-i(1-F)\pi}\cot\nu_\omega-1}-
(1-F)^2\right]+m\left[\frac{1}{e^{-i(1-F)\pi}\cot\nu_\omega-1}+1-F\right]\right\}.
\end{multline}

Taking the imaginary part of Eq.(\ref{c17}) or Eq.(\ref{c18}) at
$\omega=E+i0$, we get spectral density $\tau(E)$, see
Eq.(\ref{int9}).

%%%%%%%%%%%%%%%%%%%%%%%%%%          SECTION 4

\section{Thermal average and fluctuation of the charge}

Taking into account Eq.(\ref{int9}), one can get the following
contour integral representation for induced charge (\ref{int8})
and its quadratic fluctuation (\ref{int10}):
\begin{equation}\label{d1}
Q(T)=-\frac e2 \int\limits_C\frac{d\omega}{2\pi
i}\tanh\left(\frac12 \beta\omega\right)\,Tr(H-\omega)^{-1},
\end{equation}
and
\begin{equation}\label{d2}
\Delta^2_{Q(T)}=\frac{e^2}4 \int\limits_C\frac{d\omega}{2\pi
i}\, {\rm sech}^2 \left(\frac12
\beta\omega\right)\,Tr(H-\omega)^{-1},
\end{equation}
where $C$ is the contour $(-\infty+i0,+\infty+i0)$ and
$(+\infty-i0,-\infty-i0)$ in the complex $\omega$-plane.
Substituting Eq.(\ref{c17}) for $Re\, \omega >0$ and
Eq.(\ref{c18}) for $Re\, \omega <0$ into Eqs.(\ref{d1}) and
(\ref{d2}), we obtain
\begin{multline}\label{d3}
Q(T)=-\frac{e}2\,{\rm sgn}(m)\left\{\frac12
\left[{\rm sgn}\left(1+A^{-1}\right)-{\rm sgn}(1+A)\right]\tanh\left(\frac12
\beta|E_{BS}|\right)+\right.\\
+\frac{2\sin(F\pi)}{\pi}\int\limits_0^\infty\frac{du}{u\sqrt{u+1}}
\tanh\left(\frac12 \beta|m|\sqrt{u+1}\right)\times\\
\left.\times\frac{FAu^{F}-(1-F)A^{-1}u^{1-F}-u\cos(F\pi)+\left(F-\frac12\right)u
\left(Au^{F}+A^{-1}u^{1-F}\right)}
{\left[Au^{F}-A^{-1}u^{1-F}+2\cos(F\pi)\right]^2+4(u+1)\sin^2(F\pi)}\right\},
\end{multline}
and
\begin{multline}\label{d4}
\Delta^2_{Q(T)}=\frac{e^2}4 \left\{\frac12[1-{\rm sgn}(A)]\,
 {\rm sech}^2 \left(\frac12\beta|E_{BS}|\right)-F(1-F)\,{\rm sech}^2 \left(\frac12
\beta|m|\right)+\right.\\
+\frac{2\sin(F\pi)}{\pi}\int\limits_0^\infty\frac{du}{u}\,{\rm
sech}^2\left(\frac12 \beta|m|\sqrt{u+1}\right)\times\\
\times\left.\frac{FAu^{F}+(1-F)A^{-1}u^{1-F}-(2F-1)u\cos(F\pi)}
{\left[Au^{F}-A^{-1}u^{1-F}+2\cos(F\pi)\right]^2+4(u+1)\sin^2(F\pi)}\right\},
\end{multline}
where $A$ is given by Eq.(\ref{b18}).

In the cases of $A=0$ and $A^{-1}=0$ expressions for the charge and its
fluctuation simplify:
\begin{equation}\label{d5}
Q(T)=-\frac e2\left(F-\frac12 \pm\frac12\right)\tanh\left(\frac12 \beta
m\right), \quad \Theta=\pm s\frac \pi2\,({\rm mod}\,2\pi),
\end{equation}
and
\begin{equation}\label{d6}
\Delta^2_{Q(T)}=\frac{e^2}4\left(F-\frac12\pm\frac12\right)^2
 {\rm sech}^2 \left(\frac12\beta|m|\right), \quad \Theta=\pm s\frac \pi2\,({\rm mod}\,2\pi);
\end{equation}
note that Eq.(\ref{d5}) at ${\displaystyle\Theta= s\frac \pi2\,({\rm
mod}\,2\pi)}$ was obtained in Ref.\cite{Cor}.\vspace{0.5em}

In the limit $T\rightarrow0$ $(\beta\rightarrow\infty)$ the charge tends to
finite value (see Ref.\cite{Si7}):
\begin{equation}\label{d7}
Q(0)=\left\{
\begin{array}{lr}
\left.\begin{array}{lr}
{\displaystyle\frac e2\,{\rm sgn}(m)(1-F)},&-1<A<\infty\\
{\displaystyle\vphantom{\int}-\frac e2\,{\rm sgn}(m)F},&
A^{-1}=-1,\,\,A^{-1}=0\\ {\displaystyle\vphantom{\int}-\frac
e2\,{\rm sgn}(m)(1+F)},&-\infty<A<-1\\
\end{array}\right\},&{\displaystyle0<F<\frac12}\vspace{0.3em}\\
\left.\begin{array}{lr}
{\displaystyle\vphantom{\int}-\frac e2\,{\rm
sgn}(m)F},&-1<A^{-1}<\infty\\ {\displaystyle\vphantom{\int}\frac
e2\,{\rm sgn}(m)(1-F)},&\phantom{gfaggs}A=-1,\,\,A=0\\
{\displaystyle\vphantom{\int}\frac e2\,{\rm
sgn}(m)(2-F)},&-\infty<A^{-1}<-1\\
\end{array}\right\},&{\displaystyle\frac12<F<1}\\
\end{array}
\right.\end{equation}
\begin{equation}\label{d8}
Q(0)=
\begin{array}{lr}
\left\{\begin{array}{lr}
{\displaystyle\vphantom{\int}-\frac e\pi\, s\,{\rm
sgn}(m)\arctan\left(\tan\frac\Theta2\right)},&\Theta\neq\pi\,({\rm
mod}\,2\pi)\\ {\displaystyle0},&\Theta=\pi\,({\rm mod}\,2\pi)\\
                   \end{array}\right\}&{\displaystyle F=\frac12},\\
\end{array}\end{equation}
whereas the fluctuation tends exponentially to zero for almost all values of
$\Theta$ with the exception of one corresponding to the zero bound state
energy, $E_{BS}=0$ ($A=-1$):
\begin{equation}\label{d9}
\Delta^2_{Q(0)}=
\begin{array}{lr}
\left\{\begin{array}{lr} {\displaystyle\vphantom{\int}0,}&A\neq-1
\\ {\displaystyle\vphantom{\int}\frac{e^2}{4},}&A=-1\\
                   \end{array}\right.&.\\
\end{array}\end{equation}

In the high-temperature limit the charge tends to zero:
\begin{multline}\label{d10}
Q(T\rightarrow\infty) =\left\{\begin{array}{lr}
{\displaystyle\frac e2\,{\rm
sgn}(m)\frac{\sin(F\pi)}{\pi}\frac{\Gamma(1-F)}{\Gamma(1+F)}
\tan\left(s\frac \Theta2+\frac\pi4\right)\left(\frac{|m|}{k_BT}\right)^{1-2F}},
&{\displaystyle0<F<\frac12}\\
{\displaystyle\vphantom{\int}-\frac
e8 \frac{sm}{k_BT}\sin
\Theta},&{\displaystyle F=\frac12}\\
{\displaystyle-\frac e2\,{\rm
sgn}(m)\frac{\sin(F\pi)}{\pi}\frac{\Gamma(F)}{\Gamma(2-F)}
\cot\left(s\frac \Theta2+\frac\pi4\right)\left(\frac{|m|}{k_BT}\right)^{2F-1}}
,&{\displaystyle\frac12<F<1}\\
\end{array}\right.
\end{multline}
whereas the fluctuation tends to finite value, see Appendix B:
\begin{equation}\label{d11}
\lim_{T\rightarrow\infty}\Delta^2_{Q(T)}
=\left\{
\begin{array}{lr}
\left.\begin{array}{lr}
{\displaystyle\vphantom{\int}\!\!\!\frac{e^2}4(1-F)^2},&
{\displaystyle\Theta\neq
s\frac\pi2\,({\rm mod}\,2\pi)}\vspace{0.3em}\\
{\displaystyle\vphantom{\int}\!\!\!\frac{e^2}4F^2},& {\displaystyle\Theta=
s\frac\pi2\,({\rm mod}\,2\pi)}\vspace{0.3em}\phantom{ss\!}\\
\end{array}\right\},&{\displaystyle  0<F\leq\frac12}\vspace{0.3em}\\
\left.\begin{array}{lr} {\displaystyle\vphantom{\int}\!\!\!\frac{e^2}4F^2},&
{\displaystyle\Theta\neq
-s\frac\pi2\,({\rm mod}\,2\pi)}\vspace{0.3em}\\
{\displaystyle\vphantom{\int}\!\!\!\frac{e^2}4(1-F)^2},& {\displaystyle\Theta=
-s\frac\pi2\,({\rm mod}\,2\pi)}\vspace{0.3em}\\
\end{array}\right\},&{\displaystyle  \frac12\leq F<1}\\
\end{array}
\right.
\end{equation}

At half-integer values of $e\Phi$ one has
\begin{equation}\label{d12}
  \left.A\right|_{F=\frac12}=\tan\left(s\frac\Theta2+\frac\pi4\right),
\end{equation}
and the charge and its fluctuation take form
\begin{multline}\label{d13}
 \left.Q(T)\right|_{F=\frac12}=-\frac e4s\left\{[1-{\rm sgn}(\cos\Theta)]
 \tanh\left(\frac12\beta m\sin\Theta\right)+\right.\\
\left.+\frac{\sin2\Theta}{2\pi}\int\limits_1^\infty\frac{dv}{\sqrt{v(v-1)}}
\frac{\tanh\left({\displaystyle\frac12\beta m\sqrt{v}}\right)}{v-\sin^2{\Theta}}\right\},
\end{multline}
and
\begin{multline}\label{d14}
 \left.\Delta^2_{Q(T)}\right|_{F=\frac12}=\frac {e^2}8\left\{[1-{\rm sgn}(\cos\Theta)]\,
 {\rm sech}^2\left(\frac12\beta| m\sin\Theta|\right)-
 \frac12\,{\rm sech}^2\left(\frac12\beta| m|\right)+\right.\\
\left.+\frac{\cos\Theta}{\pi}\int\limits_1^\infty\frac{dv}{\sqrt{v-1}}\,
\frac{{\rm sech}^2\left({\displaystyle\frac12\beta |m|\sqrt{v}}\right)}{v-\sin^2{\Theta}}\right\}.
\end{multline}

An alternative representation for the charge and its fluctuation
is obtained by deforming contour $C$ to encircle poles of the
${\displaystyle\tanh\left(\frac12\beta \omega\right)}$ and
 ${\displaystyle{\rm sech}^2\left(\frac12\beta \omega\right)}$ functions, which occur
along the imaginary axis at the Matsubara modes
${\displaystyle\left(\omega_n=(2n+1)\frac{i\pi}\beta\right)}$, see Appendix C:
\begin{multline}\label{d15}
Q(T)=-e\,{\rm
sgn}(m)\left\{\frac12\left(F-\frac12\right)\tanh\left(\frac\pi{2\xi}\right)+\right.\\
+\frac{\xi}{\pi}\sum_{\begin{array}{c}\raisebox{0.3em}{${\scriptstyle
n\in\,\mathbb Z}$}\vspace{-0.7em}\\{\scriptstyle n\geq 0}\end{array}}\left.
\!\frac{2(2F-1)(2n+1)^2\xi^2+A[1+(2n+1)^2\xi^2]^F-A^{-1}[1+(2n+1)^2\xi^2]^{1-F}}
{[1+(2n+1)^2\xi^2]\{A[1+(2n+1)^2\xi^2]^F+2+A^{-1}[1+(2n+1)^2\xi^2]^{1-F}\}}\right\},
\end{multline}
and
\begin{multline}\label{d16}
\Delta^2_{Q(T)}=\frac{e^2}8[1-2F(1-F)]\,{\rm sech}^2\left(\frac{\pi}{2\xi}\right)+\\
+e^2\frac{\xi^2}{\pi^2}\sum_{\begin{array}{c}\raisebox{0.3em}{${\scriptstyle
n\in\,\mathbb Z}$}\vspace{-0.7em}\\{\scriptstyle n\geq
0}\end{array}}\,\frac{1}{[1+(2n+1)^2\xi^2]^2\{A[1+(2n+1)^2\xi^2]^F+2+A^{-1}[1+(2n+1)^2\xi^2]^{1-F}\}}\times\\
\times\left\{\vphantom{\frac11}(2F-1)[(2n+1)^2\xi^2-1]\{A[1+(2n+1)^2\xi^2]^F-A^{-1}[1+(2n+1)^2\xi^2]^{1-F}\}+\right.\\
+2\{1-[3-4F(1-F)](2n+1)^2\xi^2\}-4(2n+1)^2\xi^2\times\\
\left.\times\frac{(2F-1)\{A[1+(2n+1)^2\xi^2]^F-A^{-1}[1+(2n+1)^2\xi^2]^{1-F}\}-1+(2F-1)^2(2n+1)^2\xi^2}
{A[1+(2n+1)^2\xi^2]^F+2+A^{-1}[1+(2n+1)^2\xi^2]^{1-F}}\right\},
\end{multline}
where  $\xi=\pi/(\beta|m|)$.

%%%%%%%%%%%%%%%%%%%%%%%%%%          SECTION 5

\section{Discussion}

In the present paper we consider an ideal gas of twodimensional relativistic
massive electrons in the background of a static pointlike magnetic vortex. This
system at thermal equilibrium is found to acquire electric charge: its average
$Q(T)$ is given by Eq.(\ref{d3}), and its quadratic fluctuation
$\Delta^2_{Q(T)}$ is given by Eq.(\ref{d4}). The most general boundary
conditions (parametrized by the self-adjoint extension parameter $\Theta$) at
the location of the vortex are employed, and arbitrary values of the vortex
flux $\Phi$ are permitted; our results are periodic in $\Theta$ with period
$2\pi$ at fixed $\Phi$ and periodic in $\Phi$ with period $e^{-1}$ at fixed
$\Theta$ (e is the electron charge). Note that Eqs.(\ref{d3}) and (\ref{d4})
can be regarded as the Sommerfeld-Watson transforms of the infinite sum
representation, Eqs.(\ref{d15}) and (\ref{d16}). Note also that the charge is
odd and its fluctuation is even under transition to the inequivalent
representation of the Clifford algebra $(s\rightarrow-s\quad{\rm or}\quad
m\rightarrow-m)$.

Eq.(\ref{d3}) can rewritten in the form
\begin{equation}\label{e1}
Q(T)=Q(0)+\tilde Q(T),
\end{equation}
where $Q(0)$ is given by Eqs.(\ref{d7})-(\ref{d8})\cite{Si7}, and
\begin{multline}\label{e2}
\tilde Q(T)=\frac e2{\rm sgn}(m)\left\{\frac
{{\rm sgn}\left(1+A^{-1}\right)-{\rm sgn}(1+A)}
 {{\rm exp}\left(\beta|E_{BS}|\right)+1}+\frac{2F-1}{{\rm exp}(\beta|m|)+1}+\right.\\
\left. +\frac{\beta|m|}{2\pi}\int\limits_1^\infty dw\,{\rm sech}^2\left(\frac12\beta|m|w\right)
 \arctan\left[\frac{A(w^2-1)^F-A^{-1}(w^2-1)^{1-F}+2\cos(F\pi)}
 {2w\sin(F\pi)}\right]\right\};
\end{multline}
recall that $F$ is related to the fractional part of $e\Phi$ by
Eq.(\ref{b14}), $A$ is related to $\Theta$ by Eq.(\ref{b18}), and
bound state energy $E_{BS}$ is determined implicitly by
Eq.(\ref{b17}).

Our result should be compared with the result of
Refs.\cite{Niemi,Bab,Po}
\begin{equation}\label{e3}
Q(T)=-\frac {e^2}2 s \Phi\tanh\left(\frac12\beta m\right),
\end{equation}
where $\Phi$ is the flux of a magnetic field with an extensive
support, and it is implied that the region of the support is not
excluded. Thus, result (\ref{e3}) describes the direct effect of
the field strength, whereas our result describes the indirect,
through the vector potential, effect of the field strength from
the excluded region. In contrast to Eq.(\ref{e3}), our expressions
for
$Q(T)$ and
$\Delta^2_{Q(T)}$ are periodic in the value of the flux, vanishing
at integer values of $e\Phi$, and this can be regarded as a
manifestation of the Bohm-Aharonov effect \cite{Aha} in quantum
field theory at nonzero temperature.

The nonvanishing of the charge quadratic fluctuation signifies that the charge
of the system is not a sharp quantum observable and has to be understood as a
thermal expectation value only. In the high-temperature limit the average
charge tends to zero (\ref{d10}) and the fluctuation tends to finite value
(\ref{d11}). In the zero-temperature limit quantities $\Delta^2_{Q(T)}$ and
$\tilde Q(T)$ tend exponentially to zero and the charge becomes a sharp quantum
observable with finite value $Q(0)$ (\ref{d7})-(\ref{d8}). However, the last
statement is true for almost all values of $\Theta$ with the exception of one
corresponding to the zero bound state energy, $E_{BS}=0$ ($A=-1$), since in
this case the zero-temperature fluctuation is nonzero, see Eq. (\ref{d9}).

At half-integer values of $e\Phi$ the average charge takes form of
Eq.(\ref{d13}) which coincides (after substituting $s$ for $2eg$, where $g$ is
the magnetic monopole charge, $2eg=n$ is the Dirac quantization condition) with
the expression for the thermally induced charge in the monopole background in
threedimensional space \cite{Cor, Gol, Du}. It should be emphasized that at
non-half-integer values of $e\Phi$ the behaviour of the charge as a function of
$\Theta$ differs drastically from the one at half-integer $e\Phi$.

To see this explicitly, we plot the charge and its fluctuation as  functions of
$\Theta$ for several values of the vortex flux and temperature on
Figs.\ref{01}-\ref{09}: $F=0.1,\,\,0.3,\,\,0.5,\,\,0.7,\,\,0.9$. Here values
$(k_BT/|m|)=5^{-1},\,\,1,\,\,5$
 correspond to two dashed (with longer and shorter dashes)
 and one dotted lines, and values $T=0$ and $T=\infty$ correspond to solid lines;
 the latter cannot lead to confusion, since, as it has been already noted,
 the charge at $T=\infty$ equals to zero everywhere, while the fluctuation at $T=0$
equals to zero almost everywhere with the exception of one point
$(A=-1)$. Two dashed lines coincide practically in the utmost left parts of
 Figs.$1a$,\,$2a$ and in the utmost right parts of Figs.$4a$,\,$5a$.
 The qualitative difference between the
$F=1/2$ and $F\neq1/2$ cases is most evident at zero temperature and is
persisting with the increase of temperature, notwithstanding the dying of the
charge on the whole at high temperature.

In the ${\displaystyle F\neq\frac12}$ case the charge at zero temperature is
given by a step function with two jumps. As temperature increases, the jump
corresponding to the zero bound state energy $(A=-1)$ is smoothed out, while
another jump is persisting. The charge at $A=-1$ is not a sharp quantum
observable even at zero temperature, which is explicated by the nonvanishing of
the fluctuation in this case. As temperature departs from zero, the fluctuation
develops a maximum at $A=-1$ and a minimum close to the position of the
persisting jump of the charge, but out of the region where bound state exists.
With the increase of temperature the maximum is widening and disappearing,
while the minimum is narrowing with its position approaching the position of
the charge jump and its width  tending to zero in the high-temperature limit.

In the ${\displaystyle F=\frac12}$ case the charge at zero temperature is
linear in $\Theta$ with one jump at $A=-1$ $(\Theta=s\pi\,({\rm mod}\,2\pi))$
where the charge is not a sharp quantum observable. As temperature increases,
this jump is smoothed out. Appropriately, the fluctuation is symmetric with
respect to the position of this jump, and a maximum of the fluctuation is
smoothed out with the increase of temperature.

In conclusion we note that the system considered can acquire, in
addition to the charge, also other quantum numbers. In the case of
zero temperature this issue is comprehensively elucidated in
Refs.\cite{Si7,Si9a,Si9b}, and an appropriate generalization to
the case of nonzero temperature will be studied elsewhere.

\section*{Acknowledgements}
This work was partially supported by the State Foundation for
Basic Research of Ukraine (grant 2.7/00152), the Swiss National
Science Foundation (grant SCOPES 2000-2003 7 IP 62607) and INTAS
(grant INTAS OPEN 00-00055).

%%%%%%%%%%%%%%%%%%        APPENDIX A

{ \setcounter{equation}{0}
\renewcommand{\theequation}{A.\arabic{equation}}
\appendix
\section*{Appendix A}

The diagonal elements of $G^\omega(r,\varphi;r',\varphi')$
(\ref{c2}) satisfy second-order equations:
\begin{equation}\label{ap1}
\left[-r^{-1}\partial_rr\partial_r+r^{-2}(n-e\Phi)^2-\omega^2+m^2\right]a_n(r;r')=
(\omega+m)\frac{\delta(r-r')}{\sqrt{rr'}},
\end{equation}
\begin{equation}\label{ap2}
\left[-r^{-1}\partial_rr\partial_r+r^{-2}(n-e\Phi+s)^2-\omega^2+m^2\right]c_n(r;r')=
(\omega-m)\frac{\delta(r-r')}{\sqrt{rr'}}.
\end{equation}
The general solution to, say, Eq.(\ref{ap1}) has the form
\begin{multline}\label{ap3}
a_n(r;r')=\frac{i\pi}4(\omega+m)\left\{\theta(r-r')\left[H^{(1)}_{s(n-e\Phi)}(kr)
H^{(2)}_{s(n-e\Phi)}(kr')-\right.\right.\\
\left.\left.-H^{(2)}_{s(n-e\Phi)}(kr)H^{(1)}_{s(n-e\Phi)}(kr')\right]+H^{(1)}_{s(n-e\Phi)}(kr)\rho_n^{(a)}(r')
+H^{(2)}_{s(n-e\Phi)}(kr)\tilde\rho_n^{(a)}(r')\right\},
\end{multline}
where
\begin{equation*}
H^{(1)}_{\lambda}(u)=\frac{i}{\sin(\lambda\pi)}\left[e^{-i\lambda\pi}J_\lambda(u)-J_{-\lambda}(u)\right]
\quad{\rm and}\quad H^{(2)}_{\lambda}(u)=\frac{i}{\sin(\lambda\pi)}\left[J_{-\lambda}(u)-e^{i\lambda\pi}J_\lambda(u)\right]
\end{equation*}
are the first- and second-kind Hankel functions of order
$\lambda$. Without a loss of generality one can choose a physical
sheet for square root $k=\sqrt{\omega^2-m^2}$ as $0<{\rm
Arg}\,\,k<\pi$ $(Im \,k >0)$. Then we impose the condition that
solution (\ref{ap3}) behaves asymptotically (at
$r\rightarrow\infty$) as an outgoing wave ${\displaystyle\left(\frac{{\rm
exp}(ikr)}{2\pi\sqrt{r}}\right)}$, and this yields:
$\tilde\rho_n^{(a)}(r')=H^{(1)}_{s(n-e\Phi)}(kr')$. Thus we get
\begin{multline}\label{ap4}
a_n(r;r')=\frac{i\pi}4(\omega+m)\left\{\left[\theta(r-r')H^{(1)}_{s(n-e\Phi)}(kr)
H^{(2)}_{s(n-e\Phi)}(kr')+\right.\right.\\
\left.\left.+\theta(r'-r)H^{(2)}_{s(n-e\Phi)}(kr)H^{(1)}_{s(n-e\Phi)}(kr')\right]
+H^{(1)}_{s(n-e\Phi)}(kr)\rho_n^{(a)}(r')\right\}.
\end{multline}
In a similar way we get for the solution to Eq.(\ref{ap2})
\begin{multline}\label{ap5}
c_n(r;r')=\frac{i\pi}4(\omega-m)\left\{\left[\theta(r-r')H^{(1)}_{s(n-e\Phi)+1}(kr)
H^{(2)}_{s(n-e\Phi)+1}(kr')+\right.\right.\\
\left.\left.+\theta(r'-r)H^{(2)}_{s(n-e\Phi)+1}(kr)H^{(1)}_{s(n-e\Phi)+1}(kr')\right]
+H^{(1)}_{s(n-e\Phi)+1}(kr)\rho_n^{(c)}(r')\right\}.
\end{multline}
Quantities $\rho_n^{(a)}(r')$ and $\rho_n^{(c)}(r')$ are
determined by the condition at $r\rightarrow0$. As it has been
discussed in Section II, the condition of regularity at
$r\rightarrow0$ is imposed in the case of $n\neq n_0$, and this
yields:
$\rho_n^{(a)}(r')=H^{(1)}_{s(n-e\Phi)}(kr')$ and
$\rho_n^{(c)}(r')=H^{(1)}_{s(n-e\Phi)+1}(kr')$ $(n\neq n_0)$. Thus
we get
\begin{multline}\label{ap6}
a_n(r;r')=\frac{i\pi}2(\omega+m)\left[\theta(r-r')H^{(1)}_{|n-e\Phi|}(kr)
J_{|n-e\Phi|}(kr')+\right.\\
\left.+\theta(r'-r)J_{|n-e\Phi|}(kr)H^{(1)}_{|n-e\Phi|}(kr')\right],
\quad n\neq n_0,
\end{multline}
and
\begin{multline}\label{ap7}
c_n(r;r')=\frac{i\pi}2(\omega-m)\left[\theta(r-r')H^{(1)}_{|n-e\Phi+s|}(kr)
J_{|n-e\Phi+s|}(kr')+\right.\\
\left.+\theta(r'-r)J_{|n-e\Phi+s|}(kr)H^{(1)}_{|n-e\Phi+s|}(kr')\right],
\quad n\neq n_0,
\end{multline}
which gives the type 1 and the type 2 solutions
(\ref{c6})-(\ref{c9}).

In the case of $n=n_0$ the solutions to Eqs.(\ref{ap1}) and
(\ref{ap2}) are not regular at $r\rightarrow0$, but their
irregular behaviour has to be matched with the one of the
$b_{n_0}$ and $d_{n_0}$ components correspondingly, owing to
conditions (\ref{c13})-(\ref{c13b}). Using Eqs.(\ref{c4}) and
(\ref{c5}), we get
\begin{multline}\label{ap8}
b_{n_0}(r;r')=\frac{i\pi}4k\left\{\left[\theta(r-r')H^{(1)}_{1-F}(kr)
H^{(2)}_{-F}(kr')+\right.\right.\\
\left.\left.+\theta(r'-r)H^{(2)}_{1-F}(kr)H^{(1)}_{-F}(kr')\right]
+H^{(1)}_{1-F}(kr)\rho_{n_0}^{(a)}(r')\right\},
\end{multline}
\begin{multline}\label{ap9}
d_{n_0}(r;r')=\frac{i\pi}4k\left\{\left[\theta(r-r')H^{(1)}_{-F}(kr)
H^{(2)}_{1-F}(kr')+\right.\right.\\
\left.\left.+\theta(r'-r)H^{(2)}_{-F}(kr)H^{(1)}_{1-F}(kr')\right]
+H^{(1)}_{-F}(kr)\rho_{n_0}^{(c)}(r')\right\}.
\end{multline}
Substituting the pair of Eq.(\ref{ap4}) at $n=n_0$ and
Eq.(\ref{ap8}) into Eq.(\ref{c13}) and the pair of Eq.(\ref{ap5})
at $n=n_0$ and Eq.(\ref{ap9}) into Eqs.(\ref{c13b}), we determine
$\rho_{n_0}^{(a)}(r')$ and $\rho_{n_0}^{(c)}(r')$, and obtain the type
3 solution (\ref{c10})-(\ref{c11}).

In the absence of the vortex defect radial components for all $n$
are regular at $r\rightarrow0$:
\begin{multline}\label{ap10}
\left.a_n(r;r')\right|_{e\Phi=0}=\frac{i\pi}2(\omega+m)\left[\theta(r-r')H^{(1)}_{n}(kr)
J_{n}(kr')+\right.\\
\left.+\theta(r'-r)J_{n}(kr)H^{(1)}_{n}(kr')\right],
\end{multline}
\begin{multline}\label{ap11}
\left.c_n(r;r')\right|_{e\Phi=0}=\frac{i\pi}2(\omega-m)\left[\theta(r-r')H^{(1)}_{n}(kr)
J_{n}(kr')+\right.\\
\left.+\theta(r'-r)J_{n}(kr)H^{(1)}_{n}(kr')\right].
\end{multline}

%%%%%%%%%%%%%%%%%%        APPENDIX B

{ \setcounter{equation}{0}
\renewcommand{\theequation}{B.\arabic{equation}}
\appendix
\section*{Appendix B}

In the high-temperature limit Eq.(\ref{d4}) takes form
\begin{multline}\label{apb1}
\lim_{T\rightarrow\infty}\Delta^2_{Q(T)}
=\frac{e^2}4 \left\{\frac12[1-{\rm sgn}(A)] -F(1-F)+\right.\\
\left.+\frac{2\sin(F\pi)}{\pi}\int\limits_0^\infty\frac{du}{u}
\frac{FAu^{F}+(1-F)A^{-1}u^{1-F}-(2F-1)u\cos(F\pi)}
{\left[Au^{F}-A^{-1}u^{1-F}+2\cos(F\pi)\right]^2+4(u+1)\sin^2(F\pi)}\right\}.
\end{multline}
Using relation
\begin{multline}\label{apb2}
\frac{d}{du} \arctan\frac{(Au^F+A^{-1}u^{1-F})\sin(F\pi)}
{(Au^{F}-A^{-1}u^{1-F})\cos(F\pi)+2}=\\
=\frac{2\sin(F\pi)}{u}\, \frac{FAu^{F}+(1-F)A^{-1}u^{1-F}-(2F-1)u\cos(F\pi)}
{\left[Au^{F}-A^{-1}u^{1-F}+2\cos(F\pi)\right]^2+4(u+1)\sin^2(F\pi)},
\end{multline}
we get in the case of $-\infty<A<0$:
\begin{multline}\label{apb3}
\frac{2\sin(F\pi)}{\pi}\int\limits_0^\infty\frac{du}{u}
\frac{FAu^{F}+(1-F)A^{-1}u^{1-F}-(2F-1)u\cos(F\pi)}
{\left[Au^{F}-A^{-1}u^{1-F}+2\cos(F\pi)\right]^2+4(u+1)\sin^2(F\pi)}=\\
=\begin{array}{lr}
\left\{
\begin{array}{lr}
{\displaystyle\frac1\pi\left.\arctan\frac{(Au^F+A^{-1}u^{1-F})\sin(F\pi)}
{(Au^{F}-A^{-1}u^{1-F})\cos(F\pi)+2}\right|_{u=0}^{u=\infty}=-F,}&{\displaystyle0<F<\frac12}
\vspace{0.3em}\\
{\displaystyle\frac1\pi\left.\arctan\frac{(Au^F+A^{-1}u^{1-F})\sin([(1-F)\pi]}
{(-Au^{F}+A^{-1}u^{1-F})\cos[(1-F)\pi]+2}\right|_{u=0}^{u=\infty}=-1+F,}&{\displaystyle\frac12<F<1}\\
\end{array}\right.&.\\
\end{array}
\end{multline}
In the case of $0<A<\infty$ one should note that the integration extends over
both the principal and neighboring sheets of the Arctan function (here Arctan
$z=\arctan z+n\pi$, and $\tan($Arctan $z)=z)$:
\begin{multline}\label{apb4}
\frac{2\sin(F\pi)}{\pi}\int\limits_0^\infty\frac{du}{u}
\frac{FAu^{F}+(1-F)A^{-1}u^{1-F}-(2F-1)u\cos(F\pi)}
{\left[Au^{F}-A^{-1}u^{1-F}+2\cos(F\pi)\right]^2+4(u+1)\sin^2(F\pi)}=\\
=\begin{array}{lr} \left\{
\begin{array}{lr}
{\displaystyle1+\frac1\pi\left.\arctan\frac{(Au^F+A^{-1}u^{1-F})\sin(F\pi)}
{(Au^{F}-A^{-1}u^{1-F})\cos(F\pi)+2}\right|_{u=0}^{u=\infty}=1-F,}&{\displaystyle0<F<\frac12}
\vspace{0.3em}\\
{\displaystyle1+\frac1\pi\left.\arctan\frac{(Au^F+A^{-1}u^{1-F})\sin([(1-F)\pi]}
{(-Au^{F}+A^{-1}u^{1-F})\cos[(1-F)\pi]+2}\right|_{u=0}^{u=\infty}=F,}&{\displaystyle\frac12<F<1}\\
\end{array}\right.&.\\
\end{array}
\end{multline}
In the cases of $A=0$ and $A^{-1}=0$ we use Eq.(\ref{d6}), and in the case of
${\displaystyle F=\frac12}$ we use Eq.(\ref{d14}). Thus, we get Eq.(\ref{d11})
as the high-temperature limit of the fluctuation.

%%%%%%%%%%%%%%%%%%        APPENDIX C

{ \setcounter{equation}{0}
\renewcommand{\theequation}{C.\arabic{equation}}
\appendix
\section*{Appendix C}

If a function of complex variable $\omega$ has a pole of $l-$th order at
$\omega=\omega_n$, then the integral over a contour encircling this pole is
given by expression
\begin{equation*}
\oint d\omega\,f(\omega)=2\pi i\lim_{\omega\rightarrow\omega_n}
\frac1{(l-1)!}\frac{d^{l-1}}{d\omega^{l-1}}\left[f(\omega)(\omega-\omega_n)^l\right].
\end{equation*}
By deforming contour $C$ in Eqs.(\ref{d1}) and (\ref{d2}) to
encircle poles of first and second orders, we get
\begin{equation}\label{apc1}
Q(T)=-\frac{e}{\beta}\sum\limits_{n\in \,\mathbb
Z}Tr\,(H-\omega_n)^{-1},
\end{equation}
and
\begin{equation}\label{apc2}
\Delta^2_{Q(T)}=-\frac{e^2}{\beta^2}\sum\limits_{n\in \,\mathbb
Z}Tr\,(H-\omega_n)^{-2},
\end{equation}
where ${\displaystyle\omega_n=(2n+1)\frac{i\pi}{\beta}}$. Using
Eq.(\ref{c17}), we get
\begin{multline}\label{apc3}
Tr\,(H-\omega)^{-2}=\frac{1}{(\omega^2-m^2)^2}\left\{F(F\omega^2+2\omega
m+Fm^2)-\vphantom{\frac11}\right.
\\ -\left.2F\frac{(2F-1)\omega^2+2\omega
m+m^2}{e^{-iF\pi}\tan\nu_\omega+1}+\frac{[(2F-1)\omega+m]^2}
{\left[e^{-iF\pi}\tan\nu_\omega+1\right]^2}\right\}.
\end{multline}
Consequently, we obtain
\begin{multline}\label{apc4}
Tr\,(H-\omega_n)^{-1}=\frac1{m[1+(2n+1)^2\xi^2]}\left\{F+F^2i(2n+1)\xi\,{\rm
sgn}(m)-\vphantom{\frac11}\right.\\
 -\left.\frac{1+(2F-1)i(2n+1)\xi\,{\rm sgn}(m)}{1+A[1+(2n+1)^2\xi^2]^F[1+i(2n+1)\xi\,{\rm
 sgn}(m)]^{-1}}\right\},
\end{multline}
and
\begin{multline}\label{apc5}
Tr\,(H-\omega_n)^{-2}=\frac1{m^2[1+(2n+1)^2\xi^2]^2}\left\{F^2[1-(2n+1)^2\xi^2]+2F
i(2n+1)\xi\,{\rm sgn}(m)-\vphantom{\frac11}\right.\\
 -2F\frac{1-(2F-1)(2n+1)^2\xi^2+2i(2n+1)\xi\,{\rm sgn}(m)}{1+A[1+(2n+1)^2\xi^2]^F[1+i(2n+1)\xi\,{\rm
 sgn}(m)]^{-1}}+\\
 +\left.\frac{[1+(2F-1)i(2n+1)\xi\,{\rm sgn}(m)]^2}{\{1+A[1+(2n+1)^2\xi^2]^F[1+i(2n+1)\xi\,{\rm
 sgn}(m)]^{-1}\}^2}\right\},
\end{multline}
where $\xi=\pi/(\beta|m|)$. Summing over $n$, we get Eqs.(\ref{d15}) and
(\ref{d16}).

%%%%%%%%%%%%%%%%%        BIBLIOGRAPHY

%%%%%%%%%%%%%%%%%%%      FIGURES

\begin{figure}
\begin{tabular}{c}
\!\!\includegraphics[width=140mm]{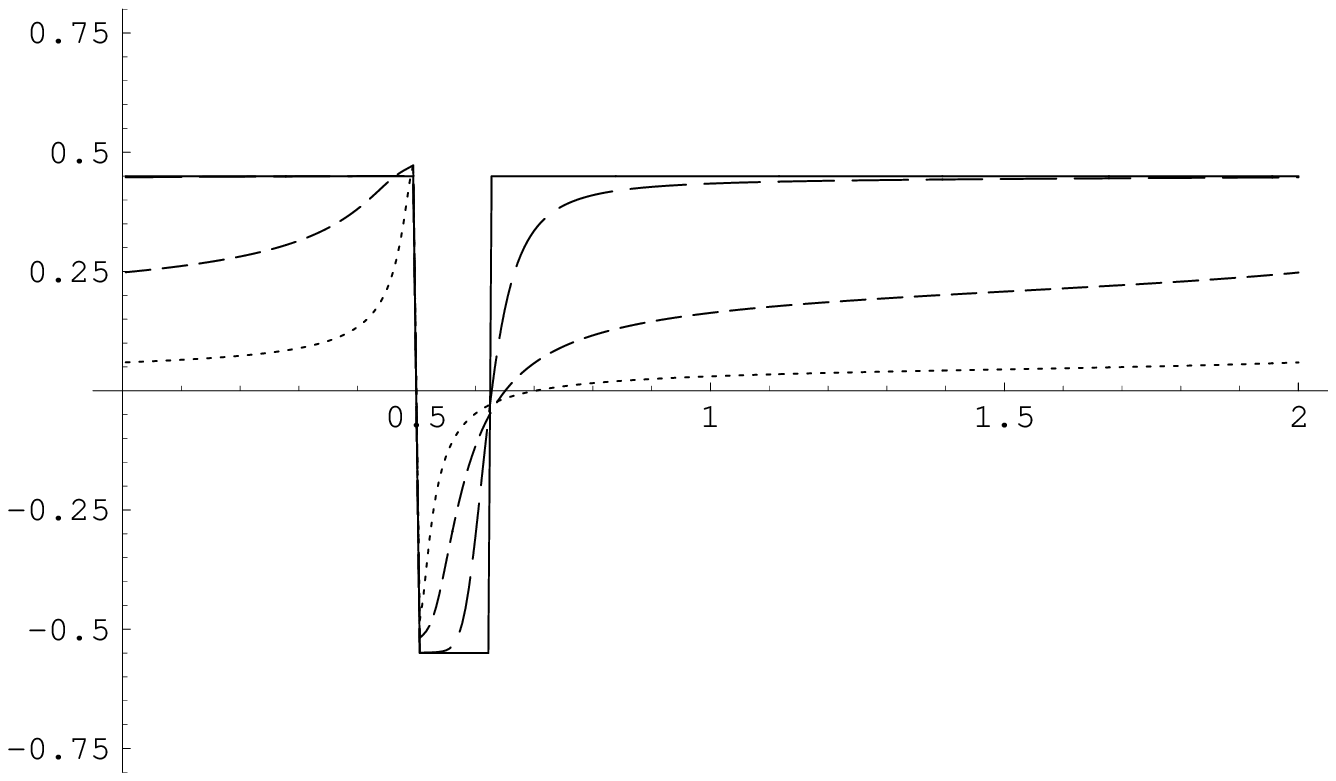}
\put(-60,220){a)}\put(-350,230){$e^{-1}Q(T)$}\put(-35,100){$s\Theta\pi^{-1}$}
\end{tabular}
\begin{tabular}{c}
\includegraphics[width=140mm]{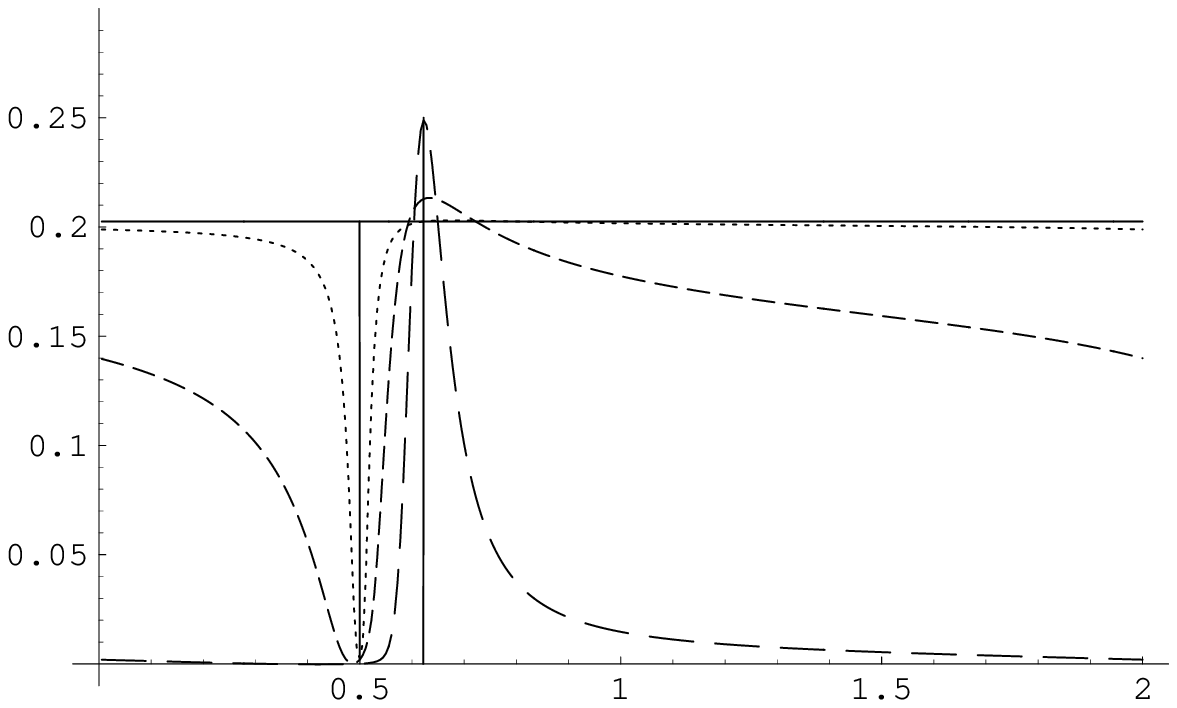}
\end{tabular}
\put(-60,110){b)}\put(-355,120){$e^{-2}\Delta^2_{Q(T)}$}\put(-40,-123){$s\Theta\pi^{-1}$}
\caption{$F=0.1$}\label{01}
\end{figure}

\clearpage
\begin{figure}
\begin{tabular}{c}
\!\!\includegraphics[width=140mm]{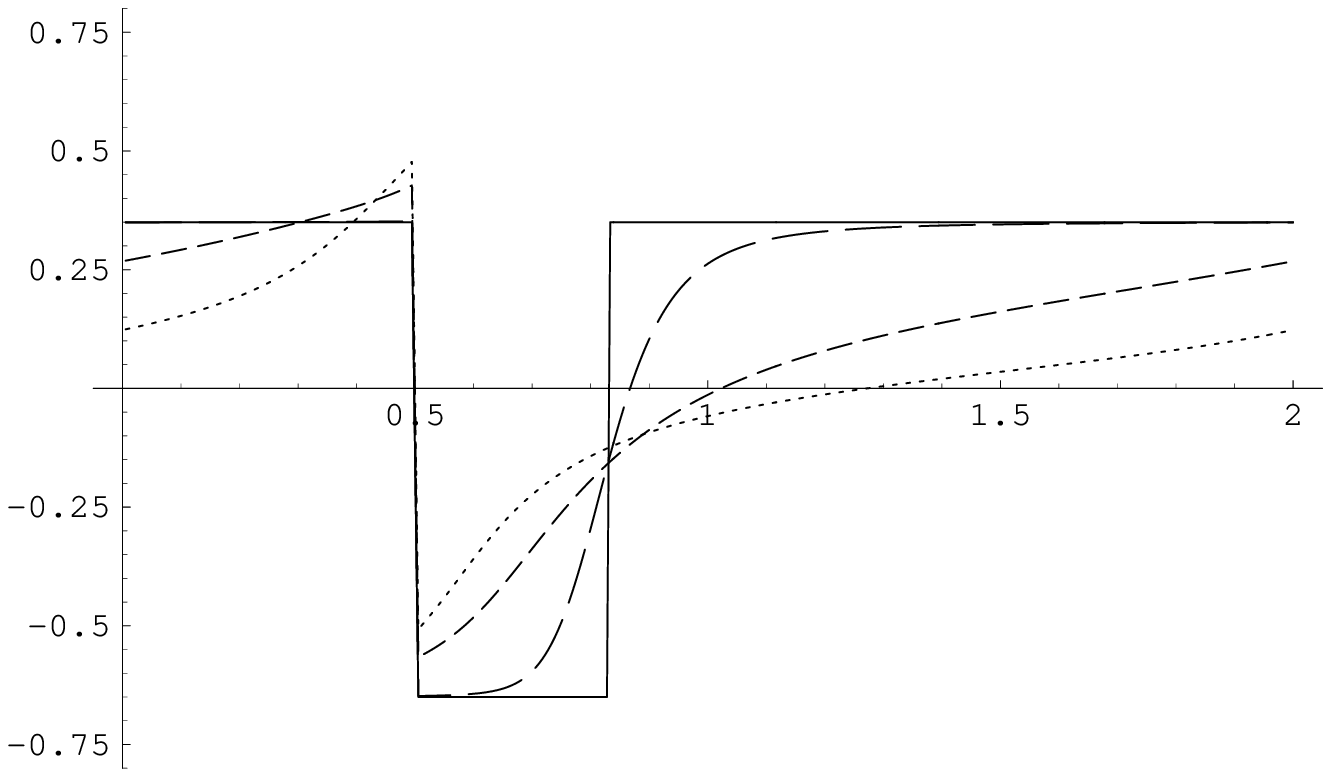}
\put(-60,220){a)}\put(-350,230){$e^{-1}Q(T)$}\put(-35,100){$s\Theta\pi^{-1}$}
\end{tabular}
\begin{tabular}{c}
\includegraphics[width=140mm]{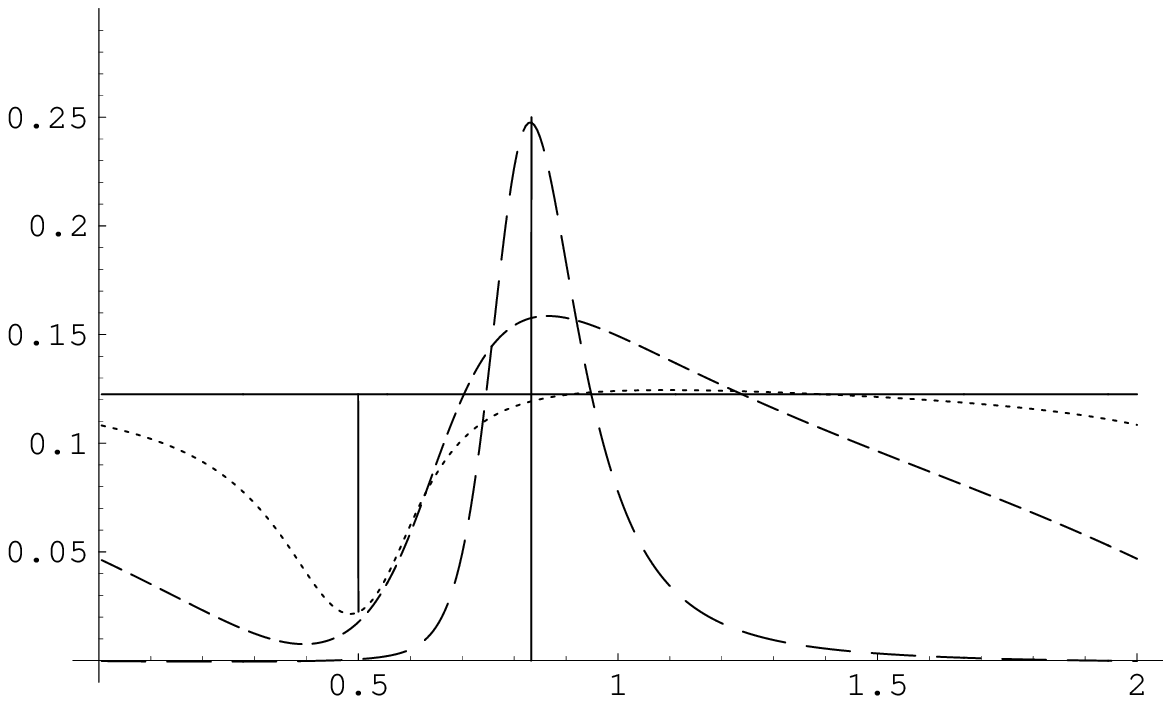}
\put(-60,180){b)}\put(-350,240){$e^{-2}\Delta^2_{Q(T)}$}\put(-34,-10){$s\Theta\pi^{-1}$}
\end{tabular}
\caption{$F=0.3$}\label{03}
\end{figure}

\clearpage
\begin{figure}
\begin{tabular}{c}
\!\!\includegraphics[width=140mm]{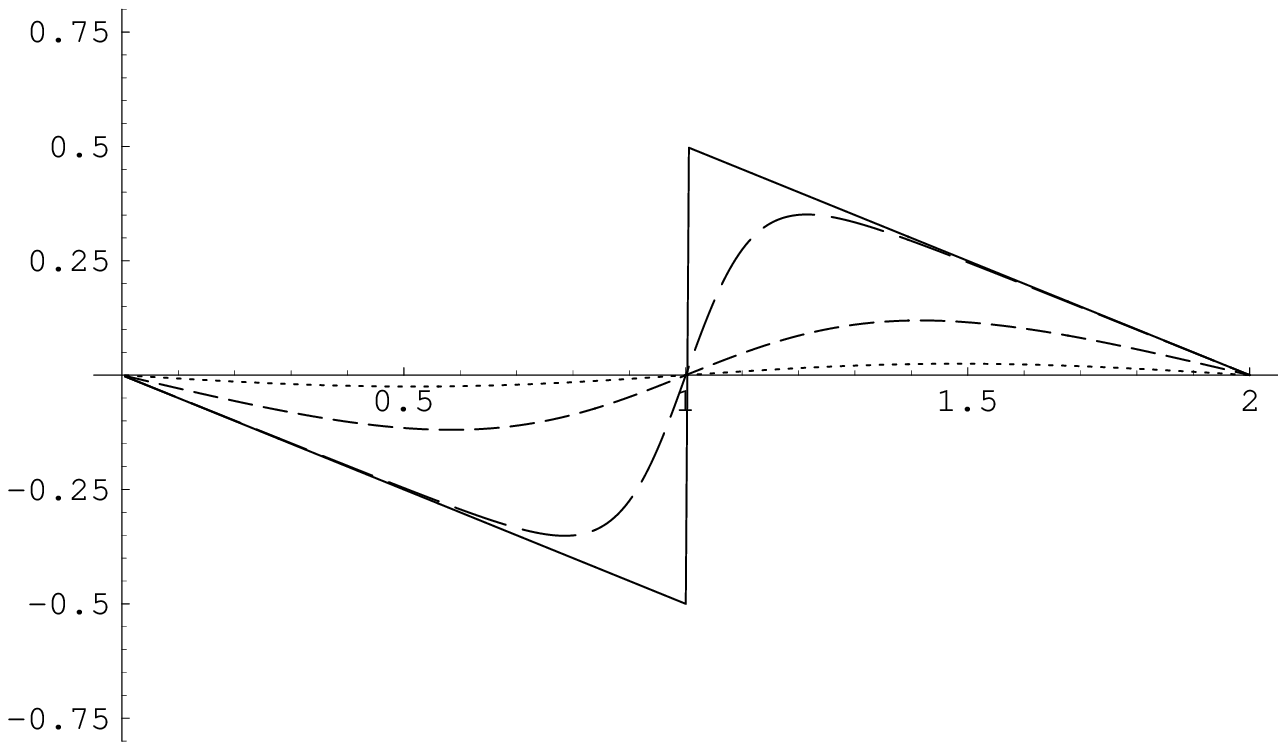}
\put(-60,220){a)}\put(-350,230){$e^{-1}Q(T)$}\put(-35,100){$s\Theta\pi^{-1}$}
\end{tabular}
\begin{tabular}{c}
\includegraphics[width=140mm]{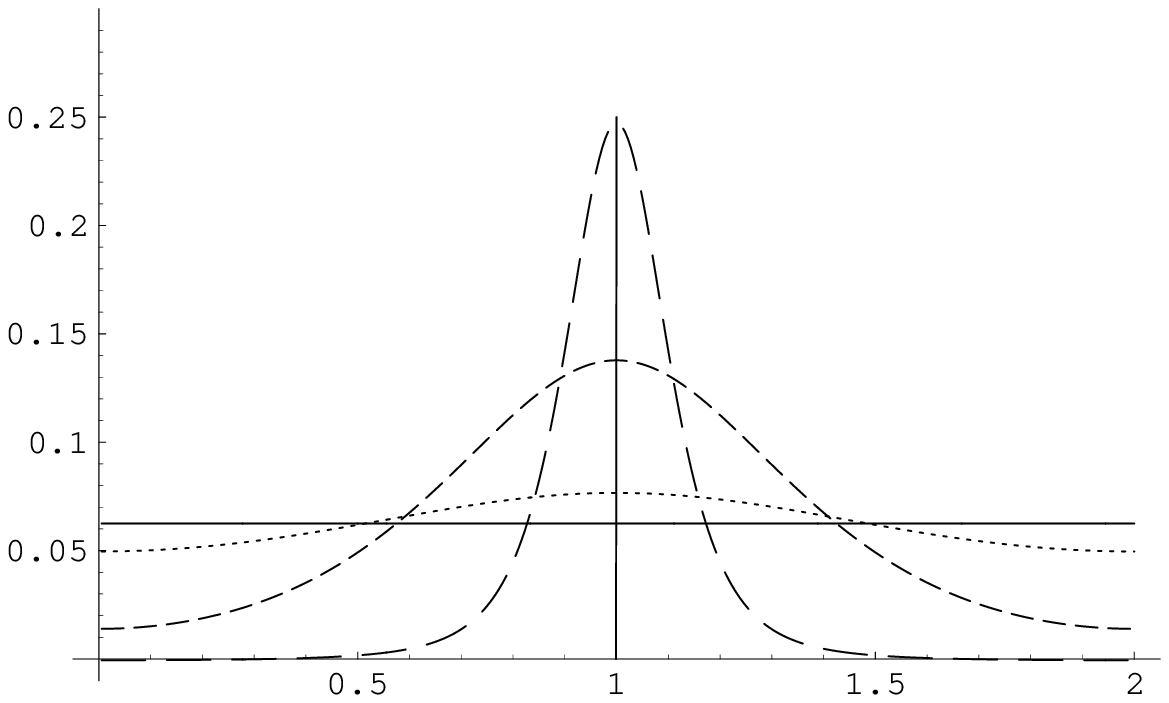}
\put(-60,200){b)}\put(-350,240){$e^{-2}\Delta^2_{Q(T)}$}\put(-34,-10){$s\Theta\pi^{-1}$}
\end{tabular}
\caption{$F=0.5$}\label{05}
\end{figure}

\clearpage
\begin{figure}
\begin{tabular}{c}
\!\!\includegraphics[width=140mm]{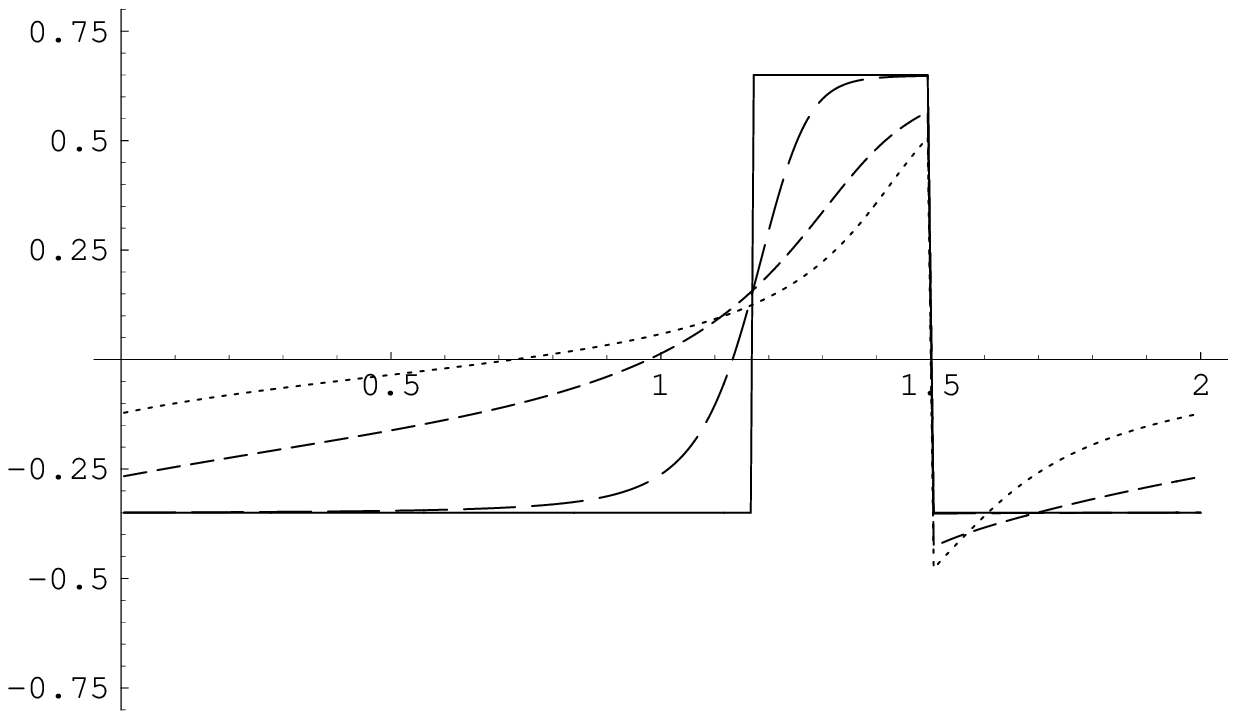}
\put(-60,220){a)}\put(-350,230){$e^{-1}Q(T)$}\put(-35,135){$s\Theta\pi^{-1}$}
\end{tabular}
\begin{tabular}{c}
\includegraphics[width=140mm]{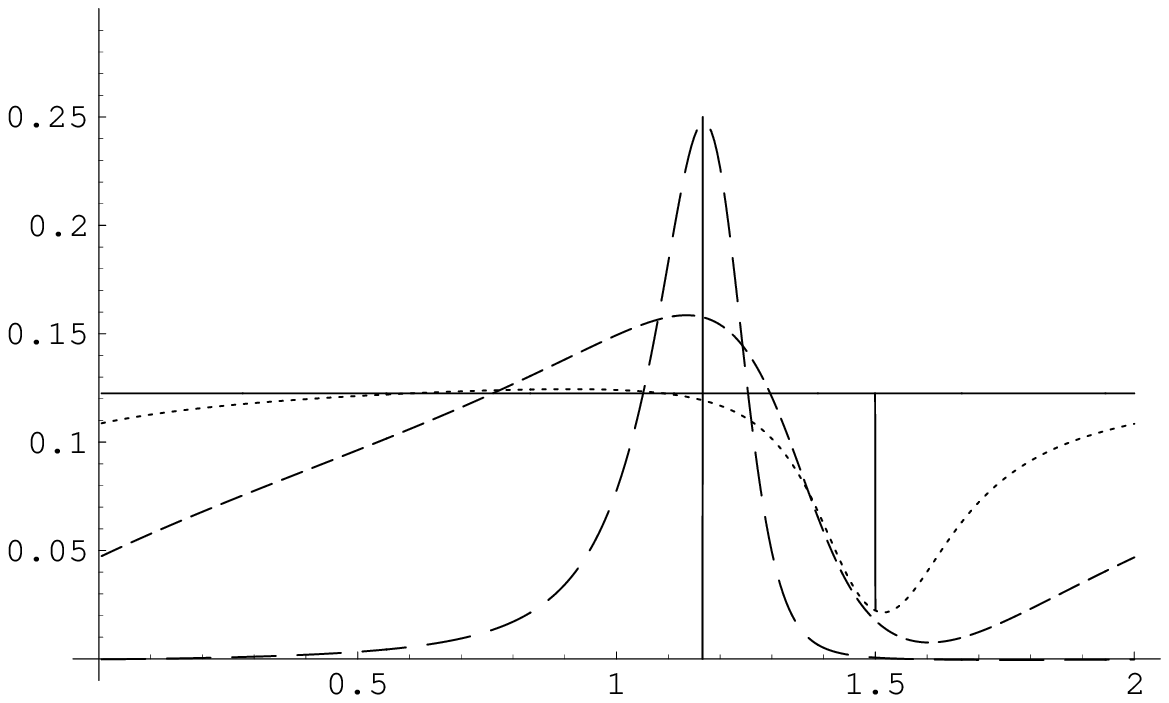}
\put(-60,200){b)}\put(-350,235){$e^{-2}\Delta^2_{Q(T)}$}\put(-34,-5){$s\Theta\pi^{-1}$}
\end{tabular}
\caption{$F=0.7$}\label{07}
\end{figure}

\clearpage
\begin{figure}
\begin{tabular}{c}
\!\!\includegraphics[width=140mm]{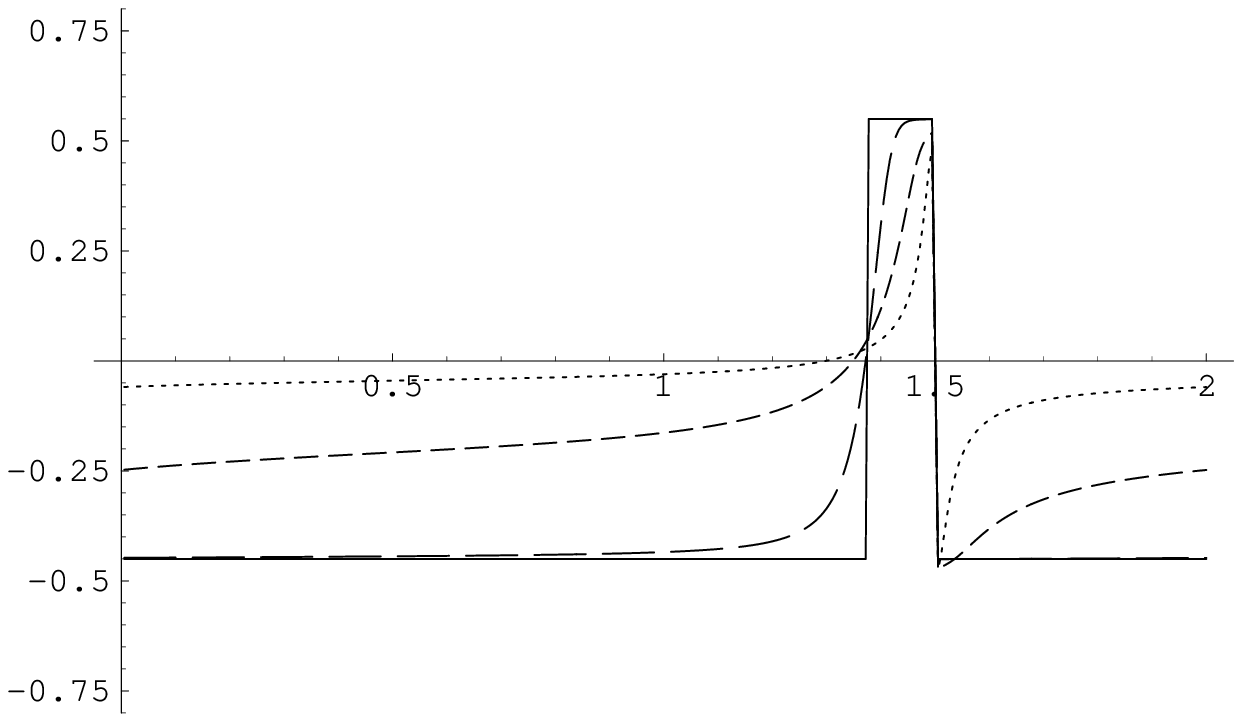}
\put(-60,220){a)}\put(-350,230){$e^{-1}Q(T)$}\put(-35,135){$s\Theta\pi^{-1}$}
\end{tabular}
\begin{tabular}{c}
\includegraphics[width=140mm]{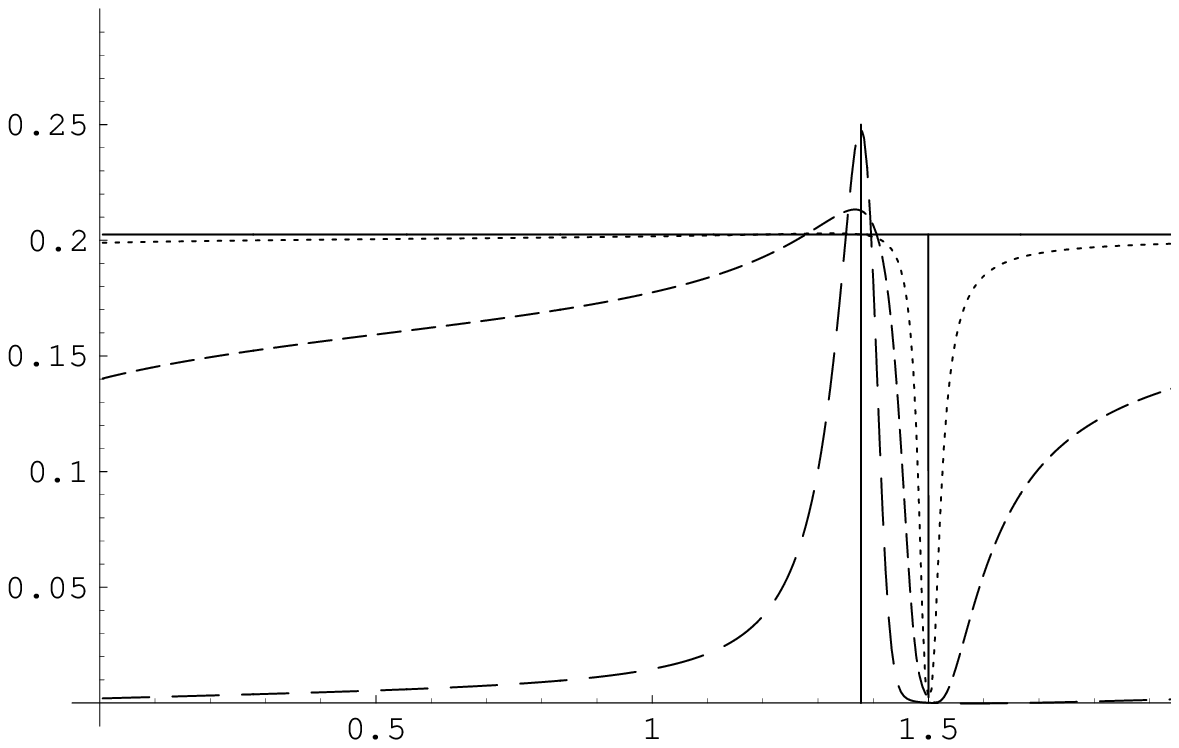}
\put(-60,210){b)}\put(-350,235){$e^{-2}\Delta^2_{Q(T)}$}\put(-34,-5){$s\Theta\pi^{-1}$}
\end{tabular}
\caption{$F=0.9$}\label{09}
\end{figure}

\end{document}